\documentclass[prd,psfig]{revtex4}
\input psfig.sty
\def\la{\mathrel{\hbox{\rlap{\hbox{\lower4pt\hbox{$\sim$}}}\hbox{$<$}}}}
\def\ga{\mathrel{\hbox{\rlap{\hbox{\lower4pt\hbox{$\sim$}}}\hbox{$>$}}}}
\def\half{{\textstyle{1\over2}}}
\def\be{\begin{equation}}
\def\ee{\end{equation}}
\def\bea{\begin{eqnarray}}
\def\eea{\end{eqnarray}}
\def\nn{\nonumber}
\def\half{{\textstyle{1\over2}}}
\newcommand{\ms}{\noalign{\vspace{3pt plus2pt minus1pt}}}

\def\bi{\bf}
\def\pd{\partial}

\def\d{d}
\def\c{{\rm c}}
\def\a{{\rm a}}
\def\<{\langle}
\def\>{\rangle}

\newfont{\myfont}{cmmib10}

\begin{document}

\title{Acceleration Mechanisms}

\author{D B Melrose}

\affiliation{School of Physics, University of Sydney, NSW 2006, Australia}

\date{\today}

\maketitle

\noindent
{\bf Article outline}

\noindent
Glossary\\
I. Background and context of the subject\\
II. Stochastic acceleration\\
III. Resonant scattering\\
IV. Diffusive shock acceleration\\
V. DSA at multiple shocks\\
VI. Applications of DSA\\
VII. Acceleration by parallel electric fields\\
VIII. Other acceleration mechanisms\\
IX. Future directions\\
X. Appendix: Quasilinear equations\\
XI. Bibliography

\medskip
\noindent
{\bf Glossary}

\noindent
{\bf Acceleration mechanism}: Any process, or sequence of processes, that results in a small fraction of the particles in a plasma each gaining an energy greatly in excess of the thermal energy.

\noindent
{\bf Heliosphere}: The region around the Sun populated by plasma escaping from the Sun in the solar wind, and extending to well beyond the orbits of the planets.

\noindent
{\bf Magnetosphere}: The region around a planet or star populated by plasma on magnetic field lines tied to the planet or star.

\noindent
{\bf Resonance}: The gyroresonance condition (\ref{gyro}) corresponds to the wave frequency being an integral, $s$, multiple  of the particle gyrofrequency in the rest frame of the particle gyrocenter; $s=0$ is also called the Cerenkov condition.

\noindent
{\bf Shock}: A fast magnetoacoustic shock is a discontinuity in plasma flow that propagates into the upstream plasma faster than the Alfv\'en speed, compressing both the plasma and the magnetic field.

\noindent
{\bf Solar corona}: The hot ($>10^6\rm\,K$) outer atmosphere of the Sun.

\noindent
{\bf Solar flare}: An explosion in the solar corona that lead to copious acceleration of fast particles.

\section{Background and Context of the Subject}

A typical astrophysical or space plasma consists of thermal plasma components, a magnetic field, various nonthermal distributions of fast particles and various kinds of turbulence involving different kinds of waves. There is no single mechanism by which the fast particles are accelerated, and it is now recognized that the data require a rich variety of different mechanisms operating under different conditions. Three generic mechanisms are emphasized in this review: stochastic acceleration, diffusive shock acceleration and acceleration by parallel electric fields. Before discussing these, it is appropriate to summarize the observational discoveries that drove the formulation of the theoretical ideas.

The major observational discoveries in this field were almost all unexpected and initially quite baffling. The first notable discovery was that of cosmic rays (CRs), by Victor Hess in 1912. Hess measured the rate of ion production inside a hermetically sealed container in a balloon flight  to an altitude of 5300 meters, and concluded that ``The results of my observation are best explained by the assumption that a radiation of very great penetrating power enters our atmosphere from above.'' By the late 1920s CRs were identified as charged particles, and not photons, and due to their isotropy it was inferred that they come from the Galaxy. Later in the 1930s it was accepted that a component of the CRs originates from the Sun. The next major discovery followed the development of radio astronomy in the 1940s. This led to the identification of a wide class of sources radiating by a nonthermal mechanism, identified in the late 1940s as synchrotron radiation. The presence of highly relativistic electrons in synchrotron sources raises the question of how they are accelerated. Subsequent development of X-ray and gamma-ray astronomy placed severe constraints on the acceleration of the relativistic electrons in synchrotron sources. 

Space physics began in the late 1950s with the discovery of the Earth's Van Allen radiation belts. Further exploration led to the identification of energetic particle distributions in essentially all space plasmas. Observations from spacecraft provide in situ data on particles and fields, in the Earth's magnetosphere, in other planetary magnetospheres, and throughout the heliosphere. In the case of the magnetosphere, Earth-based data on precipitation of particles, notably in the auroral zones, and on radio waves originating from the magnetosphere complement the spacecraft data on particles and waves in situ.

The currently accepted idea on the origin of Galactic CRs (GCRs) is associated with supernova explosions was originally proposed by Baade and Zwicky in 1934 \cite{BZ34}. Supernovae are the only adequate source of the energy needed to replace cosmic rays, whose lifetime in the Galaxy is about $10^7\rm\,yr$ \cite{GS64}. Our present-day understanding of the acceleration mechanisms involved had its forerunners in two papers by Fermi \cite{F49,F54}. In the first of these \cite{F49}, Fermi suggested that the acceleration of GCRs is due to reflections off moving interstellar clouds. Particles gain energy in head-on collisions with clouds, and lose energy in overtaking collisions; a net average gain results from head-on collisions being more probable than overtaking collisions. This model is recognized as the first example of what is called stochastic acceleration. In the second paper \cite{F54}, Fermi proposed acceleration between `closing jaws' in which all reflections are head-on and hence energy-increasing. This idea is an essential ingredient in diffusive shock acceleration (DSA), proposed independently by various authors in 1977-8 \cite{DSA1,DSA2,DSA3,DSA4}. DSA is now widely accepted as the acceleration mechanism for GCRs and for relativistic electrons in most synchrotron sources. 

Both stochastic acceleration and DSA require that the particles be scattered efficiently in order to be accelerated. Very efficient scattering is also required to explain precipitation of particles trapped in the magnetosphere. The problems associated with the scattering and acceleration of fast particles in astrophysical plasmas essentially defined a new field of `plasma astrophysics' that emerged in the mid 1960s. The efficient scattering is attributed to resonant wave-particle interactions. 

A third generic mechanism is acceleration by an electric field, $E_\parallel$, parallel to the magnetic field. There is overwhelming evidence that such acceleration occurs, most notably in the Earth's auroral zones. The acceleration is associated with a parallel current; historically, the existence of such parallel currents in the auroral zones is perhaps the oldest idea relevant to the theory of acceleration, having been predicted by Kristian Birkeland in 1903. However, understanding such acceleration is problematical. Astrophysical plasmas are typically described using ideal magnetohydrodynamics (MHD), which requires $E_\parallel=0$, and theories involving $E_\parallel$ are necessarily outside the scope of MHD MHD. Moreover, and $E_\parallel$ accelerates particles of oppositely signed charge in opposite directions, setting up a current flow, that is opposed by inductive effects, and the resulting charge separation tends to neutralize the $E_\parallel$. Hence, despite its apparent simplicity, acceleration by $E_\parallel$ is the least understood of the three generic mechanisms.

\section{Stochastic Acceleration}

Second-order Fermi acceleration \cite{F49} is now identified as an archetypical form of stochastic acceleration, and the term Fermi acceleration is sometimes used in a generic sense to describe stochastic acceleration. 

\subsection{Historical remarks}

Fermi-type acceleration occurs when particles are reflected many times from a distribution of moving
magnetic inhomogeneities. In the original version of the mechanism Fermi \cite{F49} was concerned with GCRs bouncing off magnetized interstellar clouds. Such acceleration is more efficient for more frequent, smaller energy changes than for less frequent, larger energy changes, and this led to the recognition that a more efficient Fermi-type acceleration can result from MHD turbulence \cite{Thompson55, Kaplan56, Davis56,Parker57, ParkerTidman58}. One form of the idea is that the slow compression and rarefaction of $B$, associated with the MHD turbulence, conserves the magnetic moment, $p_\perp^2/B$, and when combined with a scattering process that maintains the isotropy of the particles, this implies a `betatron' acceleration, or `magnetic pumping' \cite{Schluter57,Berger58}. A closer analogy with Fermi acceleration is acceleration due to the small fraction of the particles that reflect from the compressions in $B$, which propagate at the Alfv\'en speed, $v_{\rm A}$. Besides the betatron-type acceleration and the reflection off moving compressions, there is also a transit acceleration \cite{Shen65} due to particles diffusing through the compressions and rarefactions. All these energy changes contribute to stochastic acceleration by MHD turbulence \cite{KulsrudFerrari71}.

The mathematical description of Fermi-type acceleration is in terms of isotropic diffusion in momentum space. In the early literature stochastic acceleration was treated using a Fokker-Planck approach that includes a term that describes a systematic acceleration and a term that describes a diffusion in energy. In was shown by Tverskoi \cite{Tverskoi67} that for Fermi's original mechanism this equation reduces to an isotropic diffusion in momentum space. The same equation is derived for acceleration by isotropic turbulence \cite{Sturrock66,HallSturrock67,KulsrudFerrari71}. However, MHD turbulence is never isotropic: it consists of a spectrum of MHD waves, in both the (non-compressive) Alfv\'en mode and the (compressive) fast magnetoacoustic mode. 

\subsection{Diffusion in momentum space}

The isotropic diffusive equation in momentum space that describes stochastic acceleration is  \cite{Tverskoi67,Tverskoi68,KulsrudFerrari71}
\be
{\partial \<f\>(p)\over\partial t}=
{1\over p^2}{\partial\over\partial p}
\left[p^2D_{pp}(p)
{\partial \over\partial p}\right]\<f\>(p),
\label{(2.1)}
\ee
where $\<f\>(p)$ is the particle distribution function, $f(p)$, averaged over pitch angle. The diffusion coefficient is 
\be
D_{pp}(p)=\nu_A{cp^2\over 4v}\left(1-{v_{\rm A}^2\over v^2}\right)^2,
\label{(2.2)}
\ee
where $\nu_A$ is the acceleration rate. The final factor in (\ref{(2.2)}) does not appear in simpler treatments of Fermi-type acceleration; this factor may be ignored only in the limit of particle speeds much greater than the Alfv\'en speed, i.e., for $v\gg v_{\rm A}$. The meaning of $\nu_A$ is most easily understood by estimating the mean rate of acceleration \cite{Tsytovich66,Melrose68}
\be
\left\<{d\varepsilon\over dt}\right\>
={1\over p^2}
{\partial \over\partial p}\left[vp^2D_{pp}(p)\right],
\label{(2.3)}
\ee
where $\varepsilon=(m^2c^4+p_\parallel^2c^2+p_\perp^2c^2)^{1/2}$ is the energy. In the limit $v\gg v_{\rm A}$ (\ref{(2.3)}) gives
\be
\left\<{d\varepsilon\over dt}\right\>
\approx \nu_Apc,
\label{(2.4)}
\ee
which reduces for highly relativistic particles to the familiar form $\<d\varepsilon/ dt\>\approx \nu_A\varepsilon$. The acceleration rate may be estimated as
\be
\nu_A={\pi\over4}\,{\bar\omega}
\left({\delta B\over B}\right)^2,
\label{(2.5)}
\ee
where $\delta B$ is the magnetic amplitude in the waves and ${\bar\omega}$ is their mean frequency. A remarkable feature of (\ref{(2.5)}) is that, although effective scattering is an essential ingredient in the theory, the acceleration rate is independent of the details of the scattering. The parameters that appear in (\ref{(2.5)}) refer only to the MHD turbulence that is causing the acceleration.

\subsection{Treatment in terms of resonant interactions}

The treatment of Fermi-type acceleration by Achterberg \cite{Achterberg81} shows that it results from a resonant damping of magnetoacoustic waves in the presence of efficient pitch-angle scattering. The general gyroresonance condition is
\be
\omega-s\Omega 
-k_\parallel
v_\parallel=0,
\label{gyro}
\ee
where $\omega$ is the wave frequency, $k_\parallel$ is the component of its wave vector parallel to ${\bi B}$, $s=0,\pm1.\ldots$ is the harmonic number, $\Omega=|q|B/\gamma m$ is the relativistic gyrofrequency of the particle with charge $q$, mass $m$, Lorentz factor $\gamma$ and velocity $v_\parallel$ parallel to ${\bi B}$. The effect of gyroresonant interactions on the particles and the waves is described by the quasilinear equations, which are written down in the Appendix. The particular resonant interaction of interest here is at the Cerenkov resonance, $s=0$. Resonance at $s=0$ is possible for waves that have a component of their electric field along ${\bf B}$, and this is the case for magnetoacoustic waves, but not for Alfv\'en waves. The resonant interaction alone would cause the particles to diffuse in $p_\parallel$, with $p_\perp$ remaining constant. This corresponds to the special case
\be
{d f({\bf p})\over d t}
={\pd\over \pd p_\parallel}
\left[D_{\parallel\parallel}({\bf p})
{\pd f({\bf p})\over\pd p_\parallel}
\right]
\label{qlpar}
\ee
of the general diffusion equation (\ref{(8.18)}). In the absence of pitch-angle scattering, the distribution of fast particles becomes increasingly anisotropic, favoring $p_\parallel$ over $p_\perp$, leading to a suppression of the acceleration. In the presence of efficient pitch-angle scattering, one averages (\ref{qlpar}) over pitch-angle, leading to the isotropic diffusion equation (\ref{(2.3)}), with $D_{pp}(p)=\half\int d\mu\mu^2D_{\parallel\parallel}({\bf p})$, where $\mu=p_\parallel/p$ is the cosine of the pitch angle. The resulting expression for $D_{pp}(p)$, given by (\ref{(2.2)}) with (\ref{(2.5)}), confirms that this resonant interaction is equivalent to Fermi acceleration.

\section{Resonant scattering}

Effective scattering is an essential ingredient in stochastic acceleration and in diffusive shock acceleration. The theory of resonant scattering was developed initially in two different contexts: the diffusion of GCRs through the ISM and the formation and stability of the Earth's Van Allen belts. 

\subsection{Qualitative description of resonant scattering}

Consider the gyroresonance condition (\ref{gyro}) for $s=\pm1$ and for waves with frequency much less than the gyrofrequency, $\omega\ll\Omega$. The resonance then requires that the phase speed, $\omega/|k_\parallel|$, of the wave be much smaller than the parallel speed, $|v_\parallel|$, of the particle, and the wave is equivalent to a static fluctuation in the magnetic field. The resonance corresponds to the wavelength, $2\pi/|k_\parallel|$, of the magnetic fluctuation being equal to the distance $2\pi|v_\parallel|/\Omega$ that the particle propagates along the field line in a gyroperiod. As a consequence, the Lorentz force, $q{\bi v}\times\delta{\bi B}$, due to the magnetic fluctuation, $\delta{\bi B}$, in the wave is in the same direction at the same phase each gyroperiod. This causes a systematic change in the momentum of the particle, without any change in energy, corresponding to a change in the pitch angle, $\alpha=\arccos(p_\parallel/p)$. The sense of the change remains the same until phase coherence is lost, resulting in relatively large, random changes in $\alpha$. As a result, the particles diffuse in pitch angle. The resonance condition for a given wave (given $\omega,k_\parallel$) defines a resonance curve in velocity ($v_\parallel$--$v_\perp$) space or in momentum ($p_\parallel$--$p_\perp$) space, which is a conic section in general and which reduces to a circle in the limit $\omega\to0$. Pitch-angle diffusion corresponds to particles diffusing around the circumference of this circle. When the assumption $\omega\ne0$ is relaxed, there is a change in the energy of the particle, with the changes in $\varepsilon$ and $p_\parallel$ in the ratio $\omega:k_\parallel$.

The resonant interaction results in the waves either growing or damping, depending on whether energy is transferred from the particles to the waves, or vice versa.  For any given distribution of particles, all the particles that can resonate with a given wave lie on the resonance curve. If the gradient in the distribution function along this curve is an increasing function of $\varepsilon$, more particles lose energy to the wave than gain energy from the wave, and the wave grows. For any anisotropic distribution, there is a choice of the signs $s=\pm1$, $k_\parallel/|k_\parallel|$ and $v_\parallel/|v_\parallel|$ that implies growth of those waves that carry away the excess momentum; the back reaction on the distribution of particles, called quasilinear relaxation, reduces the anisotropy that causes the wave growth \cite{MW70}. This implies that the resonant waves needed to scatter the particles can be generated by the anisotropic particles themselves. The relevant resonant waves are Alfv\'en and magnetoacoustic waves for energetic ions and for relativistic electrons, ion cyclotron waves for nonrelativistic ions, and whistlers for nonrelativistic electrons.

\subsection{Streaming of GCRs}

Both their estimated residence time, $\sim10^7\rm\,yr$, in the Galactic disc and their small observed anisotropy imply that GCRs diffuse slowly through the Galaxy. The net flow speed is of order the Alfv\'en speed, $v_{\rm A}\sim10^{-4}c$. This diffusion is attributed to resonant scattering. A simple model for  CRs streaming along the field lines at $v_{\rm CR}$ is
\be
f(p,\alpha)=f_0(p)\left(1+{v_{\rm CR}\over v}\cos\alpha\right),
\qquad
f_0(p)=K_{\rm CR}\left({p\over p_0}\right)^{-b},
\label{(10.1)}
\ee
where  the non-streaming part of the distribution function, $f_0(p)$ is measured to be a power law above some minimum $p_0$, with $K_{\rm CR}$ a normalization constant. Observations imply $b=4.7$ below a knee in the distribution at $p\approx10^{14}$--$10^{15}{\rm\,eV}/c$ with the spectrum steepening at higher energies; Axford \cite{A94} referred to these as GCR1 and GCR2, respectively.

The anisotropic CRs can generate the resonant waves that scatter them. Growth of Alfv\'en (A) and magnetoacoustic (M) waves may be described in terms of a negative absorption coefficient. On evaluating the absorption coefficient using the general expression (\ref{(8.17)}) and inserting numerical values for cosmic rays one finds
\be
\gamma_{\rm A,M}(\omega,\theta)=-2.7\times10^{-7}
\left(
{n_e\over1{\rm\,cm^{-3}}}
\right)^{-1/2}
\left(
{p\over p_0}
\right)^{-1.6}
\left(
{\cos\theta\over|\cos\theta|}
{v_{\rm CR}\over c}
-{b\over3}\,{v_{\rm A}\over c}
\right). 
\label{growth}
\ee
where $|\cos\theta|$ is approximated by unity. It follows from (\ref{growth}) that the waves grow in the forward streaming direction ($\cos\theta>0$ for $v_{\rm CR}>0$) for $v_{\rm CR}>(b/3)v_{\rm A}$. The scattering of the CRs by the waves generated by the streaming particles themselves causes $v_{\rm CR}$ to reduce. The rate of reduction may be evaluated using the quasilinear equation written down in the Appendix. One finds
\be
{d v_{\rm CR}\over d t}=\nu_s
\left(
v_{\rm CR}
-{\zeta-1\over2}\,
{|\cos\theta|\over cos\theta}
\,{b\over3}\,v_{\rm A}
\right), 
\label{(10.8)}
\ee 
with $\zeta=({\cal F}^+_{\rm A}-{\cal F}^-_{\rm A})/({\cal F}^+_{\rm A}+{\cal F}^-_{\rm A})$, where ${\cal F}^\pm_{\rm A}$ are the fluxes of Alfv\'en waves in the forward ($+$) and backward ($-$) streaming directions, and with the scattering frequency given by
\be
\nu_s=3
\int_{-1}^{+1}\d\cos\alpha\,
\sin^2\alpha\,
D_{\alpha\alpha}, 
\label{(10.9)}
\ee
with $D_{\alpha\alpha}$ determined by (\ref{(8.19)}). 

This theory works well for lower energy GCRs. The growth rate (\ref{growth}) is fast enough to account for growth of the waves that resonate with lower energy GCRs, and the scattering time is fast enough to reduce the streaming to close to $(b/3)v_{\rm A}$. However, the growth time increases rapidly with $p$, with an approximate one-to-one relation between the resonant wave number $k$ and the momentum of the resonant particles $kp=|q|B$. For $p\gg10^{13}{\rm\,eV}/c$ the growth time becomes ineffective, and self-generation of the resonant waves cannot account for efficient scattering. 

\subsection{Scattering of higher energy CRs}

For higher-energy GCRs, effective scattering requires an external source of MHD waves. There is a turbulent spectrum in the ISM, and the observations \cite{Armstrong95} show it to be a Kolmogorov-like spectrum
\be
W(k)\propto k^{-5/3},
\label{(10.10)}
\ee
where $W(k)$ is the energy in the waves per unit wavenumber $k$. Resonant scattering by a Kolmogorov spectrum implies a scattering frequency that decreases only slowly, $\nu_s\propto p^{-1/3}$, with increasing $p$. This scattering must dominate scattering by self-generated waves at higher $p$. It is likely that the distribution of turbulence is far from homogeneous, and that the most effective scattering occurs in localized regions of enhanced turbulence. However, how effective scattering of higher energy particles occurs is a poorly understood aspect of acceleration mechanisms.

\subsection{Nonresonant versus resonant growth}

Plasma instabilities can occur in two forms: resonant and nonresonant. A nonresonant instability involves an intrinsically growing wave, described by a complex solution of a real dispersion equation. A recent suggestion that has stimulated considerable interests involves an instability \cite{BL01,Bell04} that is a nonresonant counterpart of (\ref{growth}) \cite{Melrose05}. The potential importance of this instability is twofold. First, it can allow the lower-energy particles to generate a wide spectrum of waves including those needed scatter higher-energy particles. Second, its nonlinear development can lead to amplification of the magnetic field, which ultimately causes the instability to saturate \cite{BL01,Bell04,RKirkD06}. A similar idea has been proposed in connection with generation of magnetic fields through a Weibel instability associated with the relativistic shocks that generate gamma-ray bursts \cite{WA04}. Such nonresonant growth is an important ingredient in the currently favored acceleration mechanism for GCR1 \cite{Hillas06}.

\subsection{Formation of the Van Allen belts}

Resonant scattering was first recognized in connection with particles trapped in the Earth's radiation belts, and it became an important ingredient in understanding how the radiation belts are formed and maintained. Under quiescent conditions, ions and electrons precipitate steadily (albeit in localized bursts) from the radiation belts, implying that there is a continuous resupply of fast particles. The precipitation is interpreted in terms of pitch-angle scattering of trapped particles into the loss cone. Pitch-angle scattering of ions \cite{Wentzel61,Dragt61} involves ion cyclotron waves, and the scattering of electrons involves whistlers \cite{Dungey63,KennelPetschek66}. 

\begin{figure}[t]
\centerline{
\psfig{figure=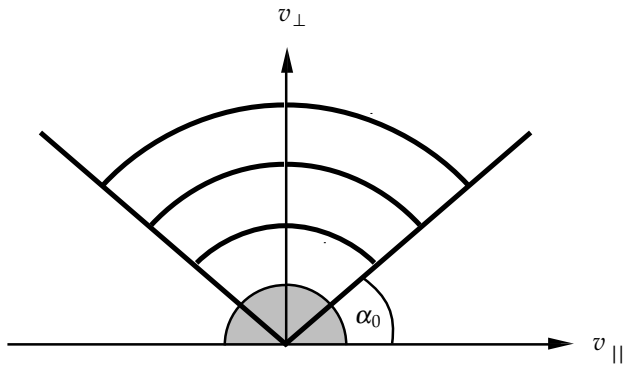,height=50mm}}
\caption{An idealized loss-cone distribution 
is illustrated; $\alpha_0$ is the loss cone angle, the shaded
region is filled by isotropic thermal particles, and the
circular arcs denote contours of constant $f$.} 
\label{fig:acc1}
\end{figure}

A simple model for the formation of the radiation belts is based on adiabatic invariants, which are associated with quasi-periodic motions. The first adiabatic invariant is associated with the gyration about a field line, and it implies that $M\propto p_\perp^2/B$ is an invariant. After averaging over the gyrations,   a fast particle bounces back and forth along a field line between two reflection points. The adiabatic invariant associated with this bounce motion, denoted $J$, is the integral of $p_\parallel$ along the field lines between the reflection points. After averaging over bounce motion, the remaining motion is a periodic drift around the Earth. There is a third adiabatic invariant, $\Phi$, which corresponds to the integral for the vector potential around the Earth. For a dipole-like magnetic field it is convenient to label a field line by the radial distance, $L$, to its midpoint. Then one has $M\propto p_\perp^2L^3$, $J\propto p_\parallel L$, $\Phi\propto1/L$. Violation of the third adiabatic invariant, due to perturbations in the Earth magnetic field with a period equal to that of the drift motion, causes the particles to diffuse in $L$ at constant $M$ and $J$. The steady state density that would result from such diffusion is proportional to $p_\perp^2p_\parallel\propto1/L^4$. Particles that enter the magnetosphere from the solar wind tend to diffuse inwards to build up this density. As the particles diffuse to smaller $L$, their momentum increases according to $p\propto L^{-3/4}$. The inward diffusion ceases at the point where pitch-angle scattering (which violates conservation of $M$) becomes effective and causes the particles to be scattered into the loss cone and to precipitate into the Earth's neutral atmosphere. Due to their loss-cone anisotropy, illustrated schematically in figure~\ref{fig:acc1}, the anisotropic particles can generate the resonant waves needed to scatter them, with the growth rate increasing rapidly with decreasing $L$. In this way, resonant scattering provides the sink (at smaller $L$) that allows the Van Allen belts to remain in a steady state when the magnetosphere is quiescent. The absorption coefficient for the waves tends to be maintained near zero, referred to as marginal stability, such that sporadic localized bursts of wave growth lead to sporadic localized bursts of particle precipitation. Such bursts of wave growth can be triggered by artificially generated VLF signals \cite{Helliwell67}.

\section{Diffusive shock acceleration}

Diffusive shock acceleration (DSA) is now recognized as the most important acceleration mechanism in astrophysical plasmas: its identification was a major achievement in the field. Prior to the identification of DSA, it was difficult to explain how acceleration can be as efficient as it appears to be, and why power-law energy spectra are so common. DSA is so efficient that one needs to look for self-regulation processes that limit its efficiency, and DSA implies power-law spectra of the form observed without any additional assumptions. 

\subsection {Description of DSA}

\begin{figure}[t]
\centerline{
\psfig{figure=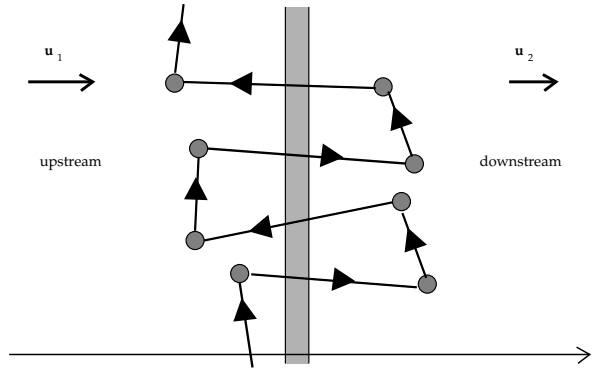,height=50mm}}
\caption{Diffusive shock acceleration is
illustrated: the shaded vertical region is the shock, the
circular blobs denote idealized scattering centers, and the
solid line with arrows denotes the path of an idealized fast
particle. The coordinate $z$ and the velocities ${\bf u}_1$
and ${\bf u}_2$ introduced in (\ref{(12.2)}) are shown for the case of a
parallel shock.} 
\label{fig:acc2}
\end{figure}

Fermi \cite{F54} suggested that the acceleration can be efficient if all reflections are head-on. A simple example is when two magnetized clouds approach each other, and a particle bounces back and forth reflecting off the approaching clouds. A related example is for a shock propagating into a closed magnetic loop, where a particle trapped in the loop can reflect head-on at the approaching shocks \cite{W64,McL71}. However, such examples are constrained by the fact that acceleration ceases when the approaching structures meet. In DSA a single shock can accelerate particles efficiently, provided that the particles are scattered on either side of the shock \cite{DSA1,DSA2,DSA3,DSA4}, as illustrated in figure~\ref{fig:acc2}.

To understand this, first note that the fluid velocity changes discontinuously across the shock. Let the two sides of the shock be labeled~1 and~2. Consider a particle on side~1 about to cross the shock and enter side~2. The scattering centers on side~2 are convected along with the plasma, and the particle sees them approaching it head on at $|u_1-u_2|$. Once the particle crosses the shock and is scattered it gains energy due to the oncoming motion of the scatterers. After being scattered a number of times on side~2, the particle can diffuse back to the shock and cross back onto side~1. On doing so, it sees the scattering centers on side~1 approaching it head on at  $|u_1-u_2|$. Again it gains energy on being scattered. DSA requires efficient scattering, and this can be achieved only by resonant scattering. Upstream of the shock the density of the fast particles decreases with distance from the shock, and the spatial gradient can cause the resonant waves to grow. Analogous to the scattering of streaming CRs, the growth of the resonant waves decreases rapidly with increasing particle energy, and some other source of resonant waves is required for higher energy particles. Scattering downstream of the shock is less problematic, with several possible sources of resonant waves, including waves generated in the upstream region and swept back across the shock.

The treatment of DSA given below is a nonrelativistic, single-particle theory. The assumption that collective effects of the fast particles can be neglected is not necessarily valid: DSA is so efficient that the fast particles can become an important component of the upstream plasma. Once the pressure associated with the fast particles becomes comparable with the thermal pressure, the structure of the shock is modified by this pressure. This can result in the stresses being transferred from the downstream to the upstream plasma primarily through the fast particles, with no discontinuity in the density of the thermal gas \cite{MD01}. Such nonlinear effects provide a constraint on DSA.

\subsection {Diffusive treatment of DSA}

Consider a distribution of particles $f(p,z)$ that is averaged over pitch angle and is a function of distance
$z$ from a shock in a frame in which the shock is at rest. It is assumed that scattering causes the particles to diffuse in the $z$~direction with diffusion coefficient $\kappa(z)$. The particles are also assumed to be streaming with the streaming speed $u$. The diffusion is described by
\bea
{\d f(p,z)\over\d t}=
{\pd\over\pd z}
\left(
\kappa(z)
{\pd f(p,z)\over\pd z}
\right) +Q(p,z) 
-f_{\rm esc}(p),
\nn
\\
\ms
{\d f(p,z)\over\d t}=
{\pd f(p,z)\over\pd t}
+u{\pd f(p,z)\over\pd z}
+{\dot p}{\pd f(p,z)\over\pd p},
\quad
{\dot p}=-{1\over3}p
{\pd u\over\pd z},
\label{(12.1)}
\eea
where $Q(p,z)$ is a source term, and where the sink term $f_{\rm esc}(p)$ takes account of escape of particles downstream from the shock. The term involving a partial derivative with respect to $p$ determines the energy changes. It is assumed that the speed $u$ changes abruptly across the shock:
\be
u=\cases{
u_1&\quad for $z<0$ \quad (upstream),
\cr
\noalign{\vskip3pt}
u_2&\quad for $z>0$ \quad (downstream),
\cr}
\qquad
{\pd u\over\pd z}=(u_1-u_2)\,\delta(z).
\label{(12.2)}
\ee

A stationary solution of (\ref{(12.1)}) exists when both the source and the sink term are neglected, such that the equation reduces to  $u\,\pd f/\pd z=(\pd/\pd z)(\kappa\,\pd f/\pd z)$; a general solution is
\be
f(p,z)=A+B\exp\left[u
\int\d z\,{1\over\kappa(z)}
\right],
\label{(12.3)}
\ee
with $u$ constant on either side of the shock. In the upstream region, $z<0$, one has $u/\kappa(z)\ne0$, and the solution (\ref{(12.3)}) diverges at $z\to-\infty$ unless one has $B=0$ for $z<0$. Writing $f_\pm(p)=\lim_{z\to\pm\infty}f(p,z)$, one has
\be
f(p,z)=\cases{
{\displaystyle
f_-(p)+[f_+(p)-f_-(p)]\exp\left[
u_1\int_0^\infty\d z\,{1\over\kappa(z)}\right]}
& for $z<0$,
\cr
\noalign{\vskip3pt}
f_+(p)&for $z>0$.
\cr}
\label{(12.4)}
\ee
On matching the discontinuity in the derivative of this solution with the discontinuity due to the
acceleration term, one finds
\be
f_+(p)=bp^{-b}\int_0^p\d p'\,
p'^{(b-1)}f_-(p'),
\quad
b={3u_1\over u_2-u_1},
\label{(12.6)}
\ee
which determines the downstream solution in terms of the upstream solution. That is, if the shock propagates into a region where the distribution is $f_-(p)$, then after the shock has passed the distribution is $f_+(p)$. 

For monoenergetic injection, $f_-(p)\propto\delta(p-p_0)$ say, (\ref{(12.6)}) implies $f_+(p)\propto p^{-b}$. The power law index, $b$, is determined by the ratio $u_1/u_2$, which is determined by the strength of the shock. In terms of the Mach number $M=u_1/v_{\rm A1}$, where $v_{\rm A1}$ is the Alfv\'en speed upstream of the shock, and the ratio $\Gamma$ of specific heats, one finds
\be
b={3r\over r-1}={3(\Gamma+1)M^2\over2(M^2-1)},
\quad
r={(\Gamma+1)M^2\over2+(\Gamma-1)M^2},
\label{(12.7)}
\ee
where $r$ is the compression factor across the shock. For $\Gamma=5/3$ and $M^2\gg1$, the
maximum strength of the shock is $r=4$ and the minimum value of the spectral index is $b=4$. 

\subsection {Alternative treatment of DSA}

An alternative treatment of DSA is as follows. Consider a cycle in which a particle with momentum $p$ crosses the shock from downstream to upstream and back again. Let $\Delta(p)$ be the change in $p$ in a cycle, and $P(p)$ be the probability per cycle of the particle escaping far downstream. Simple theory implies
\be
\Delta(p)={4(r-1)\over3r}\,{u_1\over v}\,p,
\quad
P(p)={4u_1\over vr}.
\label{(12.12)}
\ee
After one cycle, $p\to p'=p+ \Delta(p)$ implies that the number of particles per unit momentum changes according to $4\pi p^2f(p)\d p\to 4\pi p'^2f(p')\d p'=[1-P(p)]4\pi p^2f(p)\d p$. After integration, one imposes the requirement that the downstream flux of particles balances the upstream flux, and a result equivalent to (\ref{(12.6)}) is derived.

This model allows one to estimate the acceleration time, $t_\a$, in terms of the cycle time $t_\c$. One has $t_\c=(\lambda_1/u_1+\lambda_2/u_2) =(\lambda_1+r\lambda_2)/u_1$, where $\lambda_{1,2}$ are the scattering mean free paths in the upstream and downstream regions, respectively. For isotropic
scattering, the scattering mean free path, $\lambda$, is related to the spatial diffusion coefficient, $\kappa$, by $\kappa=\lambda v/3$, where $v$ is the speed of the particle. The mean free path may be written as $\lambda=\eta r_{\rm g}$, where $r_{\rm g}$ is the gyroradius, and it is usually assumed that $\eta$ is a constant of order unity, called Bohm diffusion. The rate of momentum gain is given by
\be
{\d p\over\d t}=
{p\over t_\a} 
-\left({\d p\over\d t}\right)_{\rm loss},
\quad
t_\a={u_1^2\over c\bar{\lambda}},
\quad
\bar{\lambda}={3r(\lambda_1+r\lambda_2)\over4(r-1)},
\label{(12.14)}
\ee
where $t_\a$ is an acceleration time, $\bar{\lambda}$ is a mean scattering free path, and where a loss term is included. 

\subsection{Injection and preacceleration}

DSA requires efficient scattering. For ions, scattering by Alfv\'en and magnetoacoustic waves requires $v\gg v_{\rm A}$, and for nonrelativistic electrons, scattering by whistler waves requires $v\gg43v_{\rm A}$.  These conditions are typically not satisfied for thermal particles, and one needs to postulate an injection or preacceleration mechanism to create a seed population of ions or electrons above the respective thresholds  for DSA to operate on them. 

One possible type of injection mechanism involves collisions populating a depleted thermal tail. The thresholds ($v\approx v_{\rm A}$ for ions or $v\gg43v_{\rm A}$ for electrons) are somewhere in the tail of the thermal distribution of particles, and the acceleration removes these particles, leaving a depleted thermal tail. Collisions between thermal particles cause a net flux of particles in momentum space into the high-energy tail, tending to restore a thermal distribution. The rate at which nonthermal particles are created by this process depend on the collision frequency and on the details of the acceleration mechanism \cite{Gurevich60}. However, this process encounters two serious difficulties. First, it is too slow to account for the required injection. Second, for ions it is extremely sensitive to the charge and mass, implying that the relative isotopic abundances of accelerated ions should differ by very large factors from the relative isotopic abundances of the background plasma, and this is simply not the case in general. Other preacceleration mechanisms involve either electric fields associated with the shock itself, or waves associated with the shock. These are discussed briefly below in connection with shock drift acceleration.

\subsection{DSA at relativistic shocks}

In the extension of DSA to highly relativistic shocks, the assumption of near isotropy of the pitch angle distribution is not valid, and the simple analytic theory does not generalize in a straightforward way. Numerical calculations show that the resulting spectrum is a power law, with $b=4.2$--4.3 \cite{HD89,AGKG01}.

\section{DSA at multiple shocks}

In many of the applications of DSA, it is unlikely that a single shock is responsible for the acceleration \cite{Achterberg90}. Extension of the theory to multiple shocks shows that DSA is related to Fermi-type acceleration: there are energy gains associated with DSA at each shock, and energy losses due to decompression between shocks. It is straightforward to include synchrotron losses in such a model, and the competition between acceleration and synchrotron losses can possibly explain flat or slightly inverted synchrotron spectra in some sources.

\subsection{Multiple shock model}

Consider the following model for DSA at multiple shocks.
\begin{description}
\item{(a)} All shocks are identical with compression ratio, $r=b/(b-3)$, and all have the same injection spectrum, $\phi_0(p)$, assumed to be monoenergetic, $\phi_0(p)\propto\delta(p-p_0)$.
\item{(b)} The distribution downstream of the first shock, $f_1(p)$, results from acceleration of the
injection spectrum, $\phi_0(p)$, at the shock, cf.\ (\ref{(12.6)}):
\be
f_1(p)=bp^{-b}\int_0^pdq\,q^{b-1}\phi_0(q),  
\qquad  b={3r\over
r-1}. \label{(12.15)}
\ee

\item{(c)} Between each shock a decompression occurs. Assuming that scattering keeps the particles isotropic, the distribution function after decompression is $f'_1(p)=f_1(p/R)$, with $R=r^{-1/3}$. Hence
the injection spectrum into the second shock is 
\be
f'_1(p)=b(p/R)^{-b}\int_0^{p/R}dq\,q^{b-1}\phi_0(q).
\label{(12.16)}
\ee

\item{(d)} The injection spectrum at any subsequent shock consists of the sum of  $\phi_0(p)$ and the decompressed spectrum resulting from acceleration at the previous shock.

\end{description}

\begin{figure}[t]
\centerline{
\psfig{figure=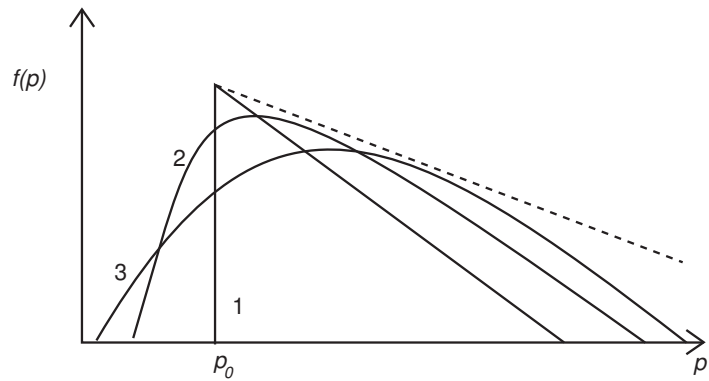,height=50mm}}
\caption{The distribution function $f'_i(p,p_0)$ for particles accelerated after $i=1$--3 shock. The dashed line indicates the asymptotic solution, $f_\infty(p)$ given by (\ref{(12.18)}).}
\label{fig:acc3}
\end{figure}

The model implies that  downstream of the $n$th shock (after decompression) the distribution is
\be
f'_n(p)=\sum_{i=1}^nf'_i(p,p_0)
\qquad
f'_i(p,p_0)={Kb^i\over p_0}\,
\left({p\over R^ip_0}\right)^{-b}\,
{(\ln\,p/R^ip_0)^{i-1}\over(i-1)!}.
\label{(12.18)}
\ee
After an arbitrarily large number of shocks, the spectrum approaches $f_\infty(p)\propto p^{-3}$ at $p>p_0$, as illustrated in figure~\ref{fig:acc3}. This spectrum is harder, by one power of $p$, than the spectrum $\propto p^{-4}$ from a single strong shock. Moreover, the spectrum $f_\infty(p)\propto p^{-3}$ is approached irrespective of the strength of the shocks, although the stronger the shocks the faster it is approached. An interpretation is that the combination of DSA and decompression leads to a Fermi-like acceleration mechanism: the asymptotic solution for Fermi-type acceleration for constant injection at $p=p_0$ is a power law with $b=3$ for $p>p_0$. Such a distribution corresponds to a flat synchrotron spectrum: the intensity of synchrotron radiation is a power law in frequency, $\nu$, with $I(\nu)\propto\nu^{-\alpha}$, $\alpha=(b-3)/2$.

\subsection{Inclusion of synchrotron losses}

\begin{figure}[t]
\centerline{
\psfig{figure=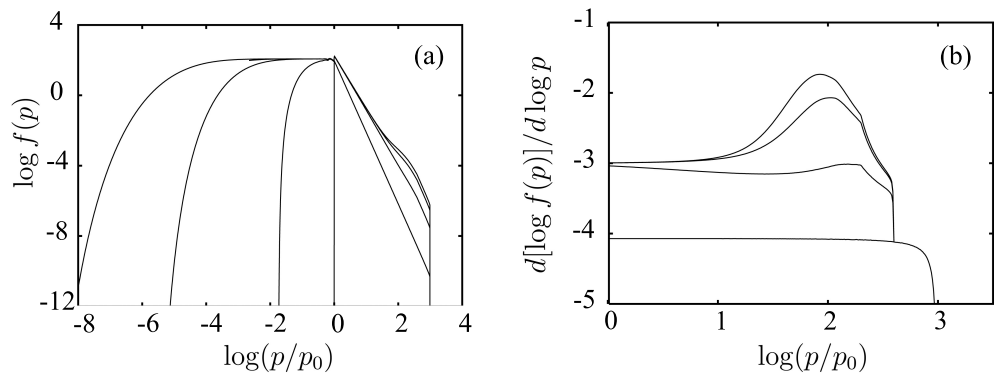,height=50mm}}
\caption{The cumulative effect of DSA plus synchrotron losses after many shocks with injection at
each shock; (a) the distribution, (b) the slope of the distribution
for $p>p_0$: $r=3.8$, $p_\c/p_0=10^3$, $N=1$, 10, 30, 50.}
\label{fig:acc4}
\end{figure}

Synchrotron losses cause the momentum of the radiating electron to decrease at the rate $dp/dt=-Ap^2$, with $A\propto B^2$. Synchrotron losses limit DSA: there is a momentum, $p=p_\c$, above which the synchrotron loss rate exceeds the rate of acceleration, and acceleration to $p>p_\c$ is not possible. The average acceleration rate over one cycle due to DSA is $dp/dt=\Delta(p)/t_\c=Cp$, and with the inclusion of synchrotron losses this is replaced by $dp/dt=Cp-Ap^2=Cp(1-p/p_\c)$ with $p_\c=C/A$. It is straightforward to repeat the calculation of DSA at multiple shocks including the synchrotron losses \cite{MC97}. In figure~\ref{fig:acc4} the logarithm of the distribution function is plotted as a function of $\log(p/p_0)$.  The synchrotron cutoff momentum is chosen to be $p_\c=10^3p_0$, and all the shocks have $r=3.8$. The distribution below the synchrotron cutoff, $p<p_\c$, due to the cumulative effect of injection at every shock, becomes harder than $b=3$, such that the slope has a peak (with $b\approx2$) just below $p_\c$. Such a distribution corresponds to a weakly inverted spectrum with a peak just below a sharp cutoff due to synchrotron losses.

\section{Applications of DSA}

There in an extensive literature on applications of DSA, and only a few remarks on some of the applications are made here. The applications may be separated into astrophysical, solar and space-physics, and the emphasis in each of these in qualitatively different. In astrophysical applications the emphasis is on the acceleration of highly relativistic particles, particularly GCRs and synchrotron-emitting electrons. In solar applications one needs to explain detailed data on the properties of the solar energetic particles (SEPs) observed in space, and on the radio, X-ray and gamma-ray signatures of them in the solar atmosphere.  In space-physics, the objective is to model the detailed distributions of particles and fields measured in situ. The comments here indicate these different emphases.

\subsection{Acceleration in young supernova remnants}

Recent observational and theoretical developments have been combined in the currently favored suggestion that GCRs are accelerated at the shocks where young supernova remnants (SNRs) interact with the interstellar medium. The observational development includes X-ray, gamma ray and TeV photon data from young supernovae, including the remnants of historical supernovae $\la10^3\rm\,yr$ old, that show edge-brightening, which is interpreted in terms of synchrotron emission from relativistic electrons accelerated at an outer shock \cite{vA04,Voelk06,Hillas06}. The suggestion is that GCRs are also accelerated at these shocks. The theoretical development is the suggestion \cite{BL01,Bell04} that the CRs can cause the magnetic field to be amplified by a very large factor in association with these shocks.

Since the original suggestion by Baade and Zwicky, it has been widely believed that the acceleration of GCRs is due to shocks generated by supernovae, and DSA provides an effective mechanism by which this occurs. Hillas \cite{Hillas06} gave four arguments in favor of the location being the shocks in young SNRs: (a) DSA produces a power-law spectrum of about the right slope. (b) The model is consistent with the long-standing arguments that the only adequate energy source for GCRs is supernovae, and that DSA is the only mechanism that is efficient enough in transferring this energy to the particles.  (c) The composition of GCR1 is consistent with ordinary interstellar matter being injected into DSA, and it is at least plausible that such composition-independent injection occurs at such shocks. (d) When coupled with self-generation of magnetic field, DSA can explain the energy at which the knee between GCR1 and GCR2 occurs \cite{Hillas06b}, cf.~however \cite{Gallant06}. Although the details are not complete, these arguments provide a plausible overview as to how GCR1 are accelerated.

The amplification of the magnetic field is attributed to the nonlinear development of a nonresonant plasma instability driven by the CRs \cite{BL01,Bell04,RKirkD06}. A strong, self-generated magnetic field allows DSA to accelerate particles to higher energy than otherwise, due to the maximum energy being proportional to $B$.

\subsection{Acceleration in astrophysical jets}

Astrophysical jets are associated with accretion discs, around proto-stars and various classes of compact objects, including the supermassive black holes in active galactic nuclei (AGN). Radio jets associated with AGN can extend to enormous distances from the AGN; these jets require that the synchrotron-emitting electrons be accelerated in situ \cite{BicknellM82}. The appearance (characterized by various knots) of the jet in M87 \cite{M87jet} in radio and optical is remarkably similar, whereas the much shorter synchrotron loss time of the optically-emitting electrons leads one to expect the optical emission to be much more sharply localized about acceleration sites. A possible model for the acceleration involves DSA at multiple shocks, which develop naturally in the flow, provided the acceleration is sufficiently fast to overcome the losses of the optically-emitting electrons. The required scattering of the optically-emitting electrons can plausibly be attributed to a Kolmogorov spectrum of turbulence  \cite{jet2}.

A subclass of radio sources associated with AGB have flat or inverted spectra, corresponding to power laws $I(\nu)\propto\nu^{\alpha}$ with $\alpha\approx0$ or $\alpha>0$ over at least a decade in frequency. Synchrotron self-absorption was proposed for such spectra \cite{K69}. A specific model can account for flat spectra \cite{flat1}, but it requires such special conditions that it was referred to as a `cosmic conspiracy' \cite{flat2}. A possible alternative explanation, for spectra with $\alpha<1/3$, is in terms of an acceleration mechanism that produces a power-law electron spectrum with $b=3-2\alpha$.  DSA at a single nonrelativistic shock produces a spectrum with $b>4$ corresponding to $\alpha<-0.5$. A relativistic shock can produce a somewhat harder spectrum but cannot account for flat spectra. DSA at multiple shocks tends towards $b=3$, $\alpha=0$, and provides a possible alternative explanation for flat spectra. The pile-up effect \cite{Schlickeiser94} occurs naturally when synchorotron losses are included in DSA at multiple shocks \cite{MC97}, and can account for inverted spectra with $\alpha\le1/3$.

\subsection{Solar energetic particles}

It has long been recognized that acceleration of fast particles occurs in the solar corona in connection with solar flares. There is a power-law distribution of flares in energy or in area: the energy released in a flare is approximately proportional to the area that brightens in H$\alpha$. All flares produce fast nonrelativistic (10--$20\rm\,keV$) electrons, observed through their radio (type~III bursts) or X-ray (bright points) emission.  Large flares also produce energetic ions and relativistic electrons. In the early literature \cite{WSW63}, it was assumed that the acceleration occurs in two phases: the first (impulsive) phase involving nonrelativistic electrons (and perhaps ions), attributed to some form of `bulk energization', and the second phase involves a slower acceleration of solar CRs, attributed to the shocks that produce type~II radio bursts. However, this simple picture was not consistent with subsequent data: gamma-ray data showed that the relativistic electrons and energetic ions appeared immediately after the onset of the flare, without the predicted delay. Spacecraft data on solar energetic particles (SEPs) revealed further phenomena and correlations that are not consistent with the two-phase model. Various different acceleration mechanisms have been explored for SEPs \cite{Miller97}.

A new paradigm has emerged for the interpretation of SEPs observed in space: events are separated into flare-associated SEPs and CME-associated SEPs \cite{Lin06}. In a CME (coronal mass ejection) a previously magnetically bound mass of corona plasma becomes detached, accelerates away from the Sun, and drives a shock ahead of it. The detachment of a CME involves some form of magnetic reconnection in the corona. Earlier ideas on a tight correlation between flares and CMEs have not been confirmed by more recent data. Although there is some correlation, it is not one-to-one, and flares and CMEs are now regarded as distinct phenomena. The acceleration of CME-associated SEPs is plausibly due to DSA, but how the flare-associated SEPs are accelerated remains poorly understood.

There are anomalies in the ionic composition in flare-associated SEPs, the most notable of which concerns ${}^3$He. The ratio of ${}^3$He to ${}^4$He in the solar photosphere is $5\times10^{-4}$, but the ratio in flare-associated SEPs is highly variable, and can be greatly enhanced, even exceeding unity in exceptional cases. The favored suggestions for preferential acceleration of ${}^3$He and other isotopic anomalies is preferential preacceleration. A specific mechanism proposed for ${}^3$He is the generation of ion cyclotron waves between the proton and ${}^4$He cyclotron resonances, with these damping at the ${}^3$He cyclotron resonance \cite{Fisk78}. The formulation of a self-consistent model for the acceleration of flare-associated SEPs is complicated by the paucity of direct signatures of processes in the solar corona, and the difficulty in reconciling the data on various emissions from energetic particles precipitating into the photosphere with the SEP data.

\begin{figure}[t]
\centerline{
\psfig{figure=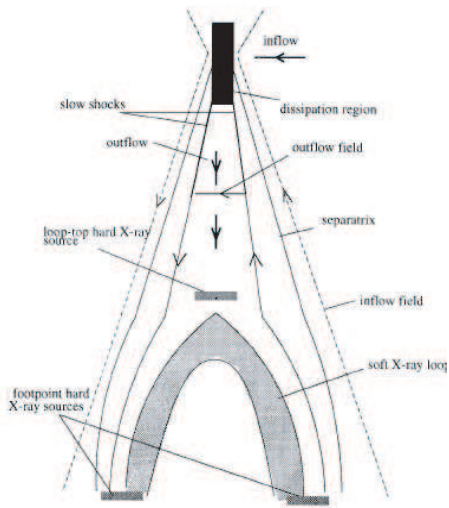,height=70mm}}
\caption{A cartoon \cite{Blackman97} illustrating one suggestion for how magnetic reconnection leads to acceleration of particles in a solar flare.}  
\label{fig:Blackman}
\end{figure}

Bulk energization of electrons is not understood. A large fraction of the energy released in a flare goes into bulk energization, requiring acceleration of virtually all the electrons in the flaring region of the corona . One model that involves shock acceleration is illustrated in figure~\ref{fig:Blackman}: magnetic reconnection leads to an outflow of plasma from the reconnection site, with the upflow associated with a CME and the downflow running into an underlying closed magnetic loop. In such a model, the electron heating is identified with shock acceleration where the downflow is stopped by the underlying loop. There are also various non-shock-associated mechanisms, involving $E_\parallel$, for bulk energization.

\section{Acceleration by parallel electric fields}

Acceleration by an electric field parallel to the magnetic field is the simplest conceivable acceleration mechanism, but also the least understood. The argument for acceleration by parallel electric fields is compelling in connection with aurorae, is very strong in connection with pulsars, and is somewhat controversial in connection with solar flares. 

\subsection{Non-MHD effects}

Acceleration by parallel electric fields, $E_\parallel$, is the accepted mechanism for auroral electrons. An essential aspect of any acceleration by $E_\parallel$ is the associated parallel current, $J_\parallel$: the relevant source term from Maxwell's equations for the energy transfer is $J_\parallel E_\parallel$. A generic model for such acceleration requires a mechanical driver, which sets up an EMF, and a closed circuit around which the current flows. The energy transport away from the driver is through a Poynting flux, propagating at $v_{\rm A}$. This energy is transferred, at least in part, to energetic particles, in an acceleration region, which acts as a load in the circuit. The load acts like a resistor, with the available potential localizing across the acceleration region, producing the $E_\parallel$. Suggested astrophysical applications of such circuit ideas tend to be viewed with suspicion, especially within the MHD community. This suspicion is partly justified due to some proposed circuit models being grossly oversimplified. However, the fact remains that there is compelling evidence for acceleration by $E_\parallel$, and the existence of $E_\parallel$ is explicitly excluded in ideal MHD. Some non-ideal-MHD effect is needed to explain why any $E_\parallel$  develops. One such mechanism is resistivity, which is included in non-ideal or resistive MHD, but the $E_\parallel$ allowed by classical resistivity is too weak to play any significant role in the fully ionized plasmas of interest. More favored possibilities include specific structures, such as double layers, or involves of Alfv\'en waves, with the $E_\parallel$ associated with inertial effects on the Alfv\'en waves.

\subsection{Available potential}

It is sometimes helpful to note that all relevant acceleration mechanisms can be attributed to an inductive electric field, and to identify the specific inductive field. There are two general approaches that provide useful semi-quantitative estimates of the available potential. One of these is the EMF associated with changing magnetic flux through a circuit; the other involves the power radiated by a rotating magnet in vacuo. 

A simple expression for the EMF is in a circuit is $\Phi=d(BA)/dt$, where $A$ is the area of the circuit. Although $\Phi$ can be due to a changing magnetic field, $\Phi=AdB/dt$, in most cases of interest it is due to a moving circuit, $\Phi=BdA/dt=BLv$, where $L$ and $v$ need to be identified in a specific model. A potential difference appears along a field line that connects two regions in relative motion to each other. The magnetic field is assumed to be frozen into both regions (MHD applies in both), and then $\Phi$ is the potential difference between them. This simple argument needs to be complemented with an identification of how the current closes, and how this potential is distributed around the closed circuit. This idea provides a plausible estimate for the energy of auroral electrons and for electrons accelerated in Jupiter's magnetosphere. In the case of auroral particles, the velocity, $v$, is between the Earth and the solar wind, and the acceleration occurs along the auroral field lines that connect these two regions. The potential available in this case can be estimated from parameters $B$ and $L$ across the magnetotail and $v$ for the solar wind, giving $\Phi$ of several kV, consistent with the observed auroral electrons of a few keV. In an early model for the Io-Jupiter interaction, the moon Io was assumed to drag the Jovian magnetic field lines threading it through the corotating magnetosphere, leading to the estimate $\Phi=B_{\rm Io}L_{\rm Io}v_{\rm Io}$, with $B_{\rm Io}$ is the Jovian magnetic field at the orbit of $Io$, $L_{\rm Io}$ is the diameter of Io and $v_{\rm Io}$ is the relative velocity of Io to the corotating magnetic field lines \cite{GL69}. This gives $\Phi$ of a few MV, which is assumed to appear around a circuit along the field lines through Io and closing across field lines in the Jovian ionosphere. 

A more subtle application of this idea provides a limit on the maximum energy to which particles can be acceleration due to DSA. The relative flow on opposite sides of the shock implies an electric field in the plane of the shock front, and one may interpret $Bv$ as this electric field with $v$ the relative flow speed perpendicular to ${\bf B}$. Then $\Phi=BvL$ is the potential available across the lateral extent, $L$, of the shock. The maximum energy to which a particle can be accelerated through DSA is limited to $<eBvL$ per unit charge. 

A point of historical interest is that the earliest recognized suggestion for acceleration of CRs involved an inductive field due to $dB/dt$: in 1933 Swann \cite{Swann33} proposed that the acceleration occurs in starspots due to the magnetic field increasing from zero to $B=0.2\rm\,T$ in a time $\tau=10^6\rm\,s$ in a region of size $L=3\times10^8\rm\,m$, giving $\Phi=BL^2/\tau=2\times10^{10}\rm\,V$. Although this model is not considered realistic, it is of interest that a similar value for $\Phi$ results from $\Phi=BLv$ with the same $B,L$ and with $L/\tau$ replaced by $v=3\rm\,km\,s^{-1}$; these numbers are close to what one would estimate for the potential difference between field lines on opposite sides of a rotating sunspot. Such large potential drops are available in the solar atmosphere, but the implications of this are poorly understood.

In a simple model for the electrodynamics of a rotating magnetized compact object, such as a neutron star or a black hole, the slowing down is estimated assuming that the rotational energy loss is due to magnetic dipole radiation. The power radiated may be written as $I\Phi$, where the effective displacement current, $I$, and the available potential, $\Phi$, are related by $\Phi=Z_0I$, with $Z_0=\mu_0c= 377\rm\,\Omega$ the impedance of the vacuum. Equating $\Phi^2/Z_0$ to the rate of rotational energy loss provides as estimate of the potential available. Although the slowing down for pulsars is not actually due to magnetic dipole radiation, this estimate remains valid for the potential available along field lines in the polar-cap region, which connect the star to the wind region outside the light cylinder. The potential may be interpreted as between the star and the light cylinder, so that it leads to an $E_\parallel$ along the polar-cap field lines that connect these regions.

\begin{figure}[t]
\centerline{
\psfig{figure=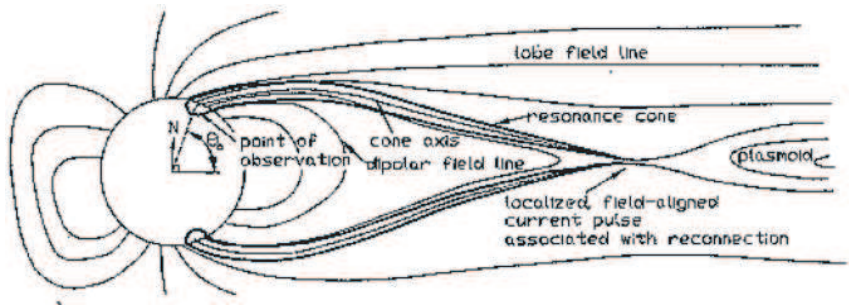,height=40mm}}
\caption{A cartoon \cite{B96} showing the regions in the Earth's magnetosphere associated with the acceleration of auroral electrons. The generation region is where magnetic reconnection occurs, launching Alfv\'en waves, which accelerate electrons in an acceleration region where the EMF from the generator is localized along field lines.}  
\label{fig:IAWs}
\end{figure}

\subsection{Acceleration of auroral electrons}

Magnetic reconnection in the Earth's magnetotail, in a so-called magnetospheric substorm, results in energetic electrons precipitating into the upper atmosphere, producing visible aurorae. A generic model involves a driver in the magnetotail, due to the solar wind sweeping field lines from the front of the magnetosphere building up the magnetic flux and associated cross-tail current, relaxation through magnetic reconnection, transport of the released energy towards the Earth in an Alfv\'enic flux at $v_{\rm A}$, and closure of the cross-tail current along field lines to the ionosphere, across the field lines due to the Pedersen conductivity and returning to the magnetotail along nearby field lines \cite{V02}. The energy flux is partially converted into energetic electrons in an acceleration region, at a height $>10^3\rm\,km$, well above the location of the visible aurora ($100\rm\,km$) and well below the magnetotail, figure~\ref{fig:IAWs}.  The acceleration is due to an $E_\parallel$ in the acceleration region, but the mechanism that sets up the $E_\parallel$ is still not clear. Three ideas are discussed briefly here.

A general mechanism that produces a parallel potential difference between the ionosphere, $\Phi_{\rm ion}$, and the magnetosphere, $\Phi_{\rm mag}$, along a given field line is associated with mirroring of thermal electrons. Let $R_{\rm m}$ be the mirror ratio, of the ionospheric to the magnetospheric magnetic fields. For $1\ll e\Phi_\parallel/T_e\ll R_{\rm m}$, the so-called Knight relation \cite{Knight73,Lyons80,VR04} gives
\be
J_\parallel\approx K(\Phi_{\rm ion}-\Phi_{\rm mag}),
\qquad
K={e^2n_e\over(2\pi)^{1/2}m_e V_e},
\label{lyons}
\ee
where $T_e=m_eV_e^2$ is the electron temperature. The current $J_\parallel$ is assumed to close across magnetic field lines due to the Pedersen conductivity of the ionosphere, leading to a relation $J_\parallel={\rm grad}_\perp(\Sigma_P{\bi E}_\perp)$, with ${\bi E}_\perp=-{\rm grad}_\perp\Phi_{\rm ion}$, where $\Sigma_P$ is the height-integrated Pedersen conductivity. However, the observed acceleration in substorms requires a more localized distribution of $E_\parallel$ in the acceleration region than is implied by the Knight relation.

In situ observations of $E_\parallel$ in the acceleration region have been interpreted in a variety of ways. Localized structures sweeping across a spacecraft have been called electrostatic shocks \cite{Mozer80} and double layers (DLs) \cite{Bostrom88,Ergun04}. It is well established in laboratory plasmas that DLs can develop, and models for DLs have been adapted to space and astrophysical plasma applications \cite{Block78,Raadu89}. DLs are classified as weak or strong, depending on whether the potential difference is of order or much greater than $T_e/e$, respectively.  The formation and structure of DLs can involve two other ideas that have been discussed widely in the literature: anomalous resistivity and electron phase space holes \cite{Newman01}. However, the details of how localized DLs accelerate particles and produce the observed fluxes of energetic electrons is still not clear.

A third way in which an $E_\parallel$ can arise is through the dispersive properties of Alfv\'en waves \cite{S00,kAw02,V02}: the condition $E_\parallel=0$ is strictly valid for an Alfv\'en wave only in ideal MHD. Inhomogeneities probably play several important roles. First, an inhomogeneity along the field lines can cause the Alfv\'en speed to change, leading to reflection of Alfv\'en waves, and localized trapping of their energy in an Alfv\'en resonator \cite{LysakS05, Ergun06}, between reflection at the top of the ionosphere and at an inhomogeneous region higher in the magnetosphere. Second, inhomogeneities across the field lines can lead to the Alfv\'en waves developing large $k_\perp$, such that the Alfv\'en waves develop an $E_\parallel$, becoming inertial Alfv\'en waves \cite{V02,kAw02}. This modification is significant for low-frequency Alfv\'en waves with large $k_\perp$: with $\lambda_e=c/\omega_p$ the skin depth, the dispersion relation and the ratio of the parallel to the perpendicular electric fields become
\be
{\omega\over|k_\parallel|}={v_{\rm A}\over(1+k_\perp^2\lambda_e^2)^{1/2}},
\qquad
{E_\parallel\over E_\perp}={k_\parallel k_\perp\lambda_e^2\over1+k_\perp^2\lambda_e^2}.
\label{IAW}
\ee
The $E_\parallel$ associated with inertial Alfv\'en waves  is a favorable mechanism for acceleration. The suggestion that the acceleration is due to the $E_\parallel$ in an inertial Alfv\'en waves is a relatively old idea \cite{GB79}, in particular for electrons accelerated by Alfv\'en waves generated by Io's motion through the Jovian magnetosphere \cite{Goertz}. This $E_\parallel$, being a wave field, reverses sign, accelerating electrons in opposite directions,  every half wave period; the wave period is assumed to be of order the Alfv\'en propagation time across the Alfv\'en resonator. The details of this acceleration are still not adequately understood \cite{Ergun02}.

\begin{figure}[t]
\centerline{
\psfig{figure=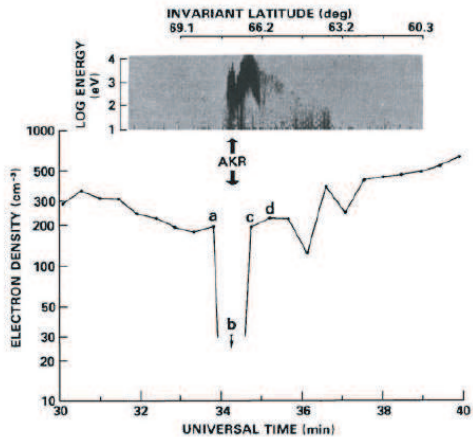,height=60mm}}
\caption{A density depletion in an auroral region of the Earth; inverted-V electrons carry a large upward current in such regions \cite{Benson}.}  
\label{fig:cavity}
\end{figure}

An extreme example of auroral acceleration occurs in inverted-V electrons. The upward current can essentially depopulate a localized auroral flux tube of all thermal electrons, leading to a plasma cavity in which the inverted-V electrons are observed, figure~\ref{fig:cavity}. The `inverted-V' refers to the dynamic spectrum observed by a spacecraft as it crosses the cavity: lowest energies near the edges and maximum energy near the center. Inverted-V electrons correlate with auroral kilometric radiation \cite{Gurnett74}. 

\subsection{Acceleration in pulsars gaps}

Pulsars are rapidly rotating neutron stars with superstrong magnetic fields. There are three broad classes: ordinary, millisecond or recycled, and magnetars; only ordinary pulsars are considered here. These have periods $P\approx0.1$--$1\rm\,s$, surface magnetic fields $B\approx10^7$--$10^9\rm\,T$ and are slowing down at a rate ${\dot P}\sim10^{-15}$. The age, $P/2{\dot P}$, and the surface magnetic 
field, $B\propto(P{\dot P})^{1/2}$, are estimated based on the model in which the rotational energy is due to magnetic dipole radiation. Equating the rate rotation energy is  lost due to the observed slowing down to $\Phi^2/Z_0$ is thought to give a valid estimate for the potential available along field lines in the polar-cap region. For the Crab pulsar the rotational energy loss is approximately $10^{31}\rm\,W$, implying $\Phi\approx6\times10^{16}\rm\,V$; the available potentials in most other pulsars are somewhat smaller than this. In the most widely favored model, the electric field is assumed to cause a charge-limited flow from the stellar surface, providing the `Goldreich-Julian' charge density required by the corotation electric field. However, the Goldreich-Julian charge density cannot be set up globally in this way, and the available potential develops across localized regions called gaps. The acceleration of `primary' particles occurs in the gaps, and gamma rays emitted by these particles trigger a pair cascade that produces a secondary (electron-positron) plasma. This additional source of charge is assumed to screen $E_\parallel$ outside the gaps. However, there is no fully consistent model and no consensus on the details of the structure and location of the gaps. The present author believes that existing models, based on a stationary distribution of plasma in a corotating frame, are violently unstable to temporal perturbations, resulting in large-amplitude oscillations in $E_\parallel$ \cite{Levinson}.

Although the details of acceleration by $E_\parallel$ are quite different in the pulsar case from the auroral case, the underlying difficulties in understanding such acceleration are similar. A static model for acceleration by $E_\parallel$ encounters seemingly insurmountable difficulties, and an oscillating model involves unfamiliar physics.

\subsection{Acceleration of nonrelativistic solar electrons}

A large fraction of the energy released in a solar flare goes into very hot electrons: all the electrons in a relatively large volume are accelerated in bulk to 10--$20\rm\,keV$. Several models for bulk heating involving acceleration by $E_\parallel$ have been proposed, but all have unsatisfactory features. One idea is to invoke a runaway process \cite{Holman85}. Consider the equation of motion of an electron in the presence of an electric field and of a frictional force due to collisions with a collision frequency $\nu_e(v)=\nu_e(V_e/v)^3$, where $V_e$ is the thermal speed of electrons, is of the form 
\be
{d{\bi v}\over dt}=-{e{\bi E}\over m}
-\nu_e{\bi v}\left({V_e\over v}\right)^3.
\label{(2.13)}
\ee
It follows that an electron with speed
\be
{v\over V_e}> \left({E_D\over E}\right)^{1/2},
\quad
E_D={mV_e\nu_e\over e},
\label{(2.14)}
\ee
are freely accelerated; $E_D$ is called the Dreicer field. However, electron runaway sets up a charge separation that should quickly screen $E_\parallel$ and suppress the acceleration. It also tends to cause $J_\parallel$, which is opposed by inductive effects. Runaway on its own is not an acceptable acceleration mechanism.

If bulk energization is due to acceleration by an $E_\parallel$, then the energies involved imply an accelerating $\Phi$ of order 10--$20\rm\,kV$. Two arguments suggest that much larger potential is not only available but also required. On the one hand, there is observational evidence that large flares tend to occur in flux loops carrying large currents, $I\ga10^{12}\rm\,A$. If one identifies $I$ with the observed current, then to produce the power $I\Phi$ of order $10^{22}\rm\,W$ released in a large flare requires $\Phi$ of order $10^{10}\rm\,V$. With $\Phi=10^5\rm\,V$, to account for a power of $10^{22}\rm\,W$ would require a current of $10^{17}\rm\,A$, which greatly exceeds the maximum possible direct current, $\sim10^{13}\rm\,A$, that can flow in a coronal flux loop. On the other hand, as indicated above, Swann's \cite{Swann33} old argument for a potential of order $10^{10}\rm\,V$ is consistent with a more plausible estimate for the available potential due to relative motion of magnetic footpoints on the solar surface. How the available potential of $10^9$--$10^{10}\rm\,V$ localizes into regions with $E_\parallel\ne0$ of length $L\approx{10^5\rm\,V}/E_\parallel$ is not explained adequately in any existing model. 

As with the auroral and pulsar cases, any model for acceleration by a static $E_\parallel$ in the solar corona encounters serious difficulties. Although a large-scale oscillating $E_\parallel$ might plausibly overcome these difficulties, there is no model for such oscillations, which would necessarily involve non-MHD effects. Modification of the ideas proposed for acceleration of auroral electrons by inertial Alfv\'en waves in an Alfv\'en resonator is perhaps a plausible starting point for a more plausible model for bulk energization.

\section{Other acceleration mechanisms}

The three most important acceleration mechanisms are DSA, stochastic acceleration and acceleration by parallel electric fields. Various other mechanisms are relevant in specific applications. Examples mentioned above are the acceleration of energetic particles in the Earth's Van Allen belts and various preacceleration mechanisms for DSA. Some comments are made here on three other mechanisms: shock drift acceleration, gyroresonant acceleration and acceleration in current sheets.

\subsection{Shock drift acceleration (SDA)}

When a particle encounters a shock it is either transmitted through the shock front or reflected from it, and in either case it tends to gain energy \cite{Toptygin80}. Shock drift acceleration (SDA) is attributed to the grad-$B$ drift when the particle encounters the abrupt change in $B$: the scalar product of drift velocity and acceleration by the convective electric field, $-q{\bi u}\times{\bi B}$ is positive, implying energy gain. The energy gained by a particle depends on how far it drifts along the front.

\begin{figure}[t]
\centerline{\psfig{figure=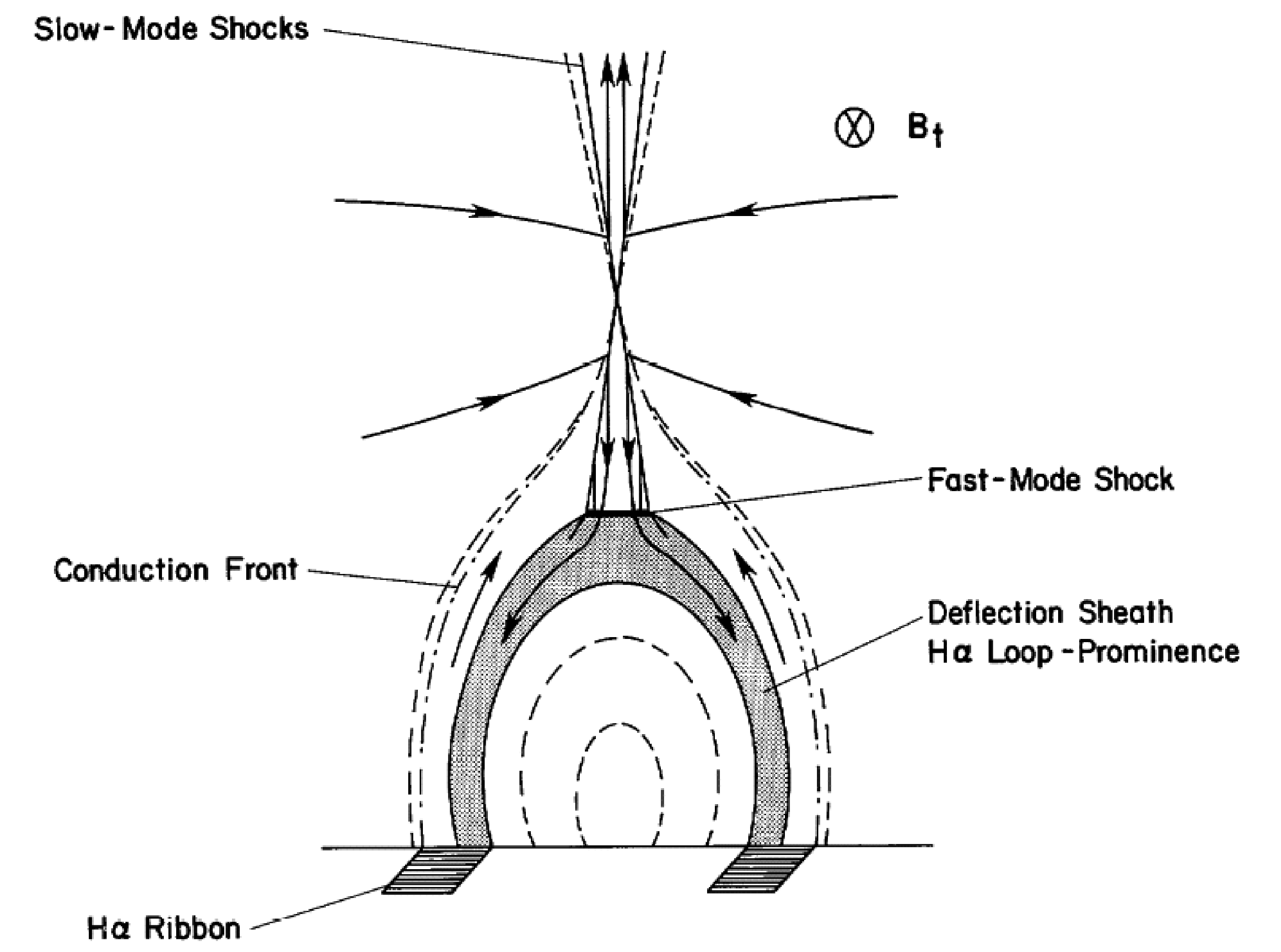,height=50mm}}
\caption{In the de Hoffmann--Teller
frame the fluid velocities are parallel to the magnetic fields on
either side of the shock.}  
\label{fig:acc5}
\end{figure}

In treating SDA it is convenient to make a Lorentz transformation to the de Hoffmann--Teller frame. The frame most widely used in discussing a shock is the shock-normal frame, in which the shock front is stationary and the upstream flow velocity is normal to it. In this frame there is a convective electric field $-{\bi u}\times{\bi B}$ on both sides of the shock, and the point of intersection of ${\bf B}$ with the shock moves in the plane of the shock. In the de Hoffmann--Teller frame this point of intersection is stationary,  ${\bi u}$ and ${\bi B}$ are parallel to each other on both sides of the shock, and there is no electric field, figure~\ref{fig:acc5}. (The de Hoffmann--Teller frame exists only if this velocity of intersection is less than $c$; in the opposite case, there exists a frame in which ${\bf B}$ is perpendicular to the shock normal.) In the de Hoffmann--Teller frame the energy of particles is unchanged on crossing or being reflected from the shock. A simple theory for SDA is based on assuming that $p_\perp^2/B$ is conserved, which is approximately the case in a gyrophase-averaged sense \cite{Pesses81,Drury83}, considering the conditions for reflection and transmission in the de Hoffmann--Teller frame, and Lorentz transforming to the shock-normal frame.

Let quantities in the de Hoffmann--Teller and shock-normal frames be denoted with and without primes, respectively. Let $\psi_1$ be the angle between ${\bf u}_1$ (the shock normal) and ${\bi B}_1$. Provided that the relative velocity between the two frames is nonrelativistic, the angle $\psi_1$ is the same in the two frames, and the relative velocity is $u_0=u_1\tan\psi_1$. In the nonrelativistic case, the pitch angles in the two frames are related by $v'\cos\alpha'=v\cos\alpha+u_0$, $v'\sin\alpha'=v\sin\alpha$. Reflection is possible only for $\sin^2\alpha'_1\ge B_1/B_2$, and the maximum change in energy occurs for reflected particles at the threshold, $\alpha'_1=\alpha_c$, $\sin^2\alpha_c=B_1/B_2$. The maximum ratio of the energy after reflection (ref) to before (inc) is for a particle with $v_1\cos\alpha_1=-u_0(1-\cos\alpha_c)$, that is propagating away from the shock, and is overtaken by the shock. The maximum ratio is
\be
\left(\frac{E_{\rm ref}}{E_{\rm inc}}\right)_{\rm max}=\frac{1+(1-B_1/B_2)^{1/2}}{1-(1-B_1/B_2)^{1/2}},
\label{maxratio}
\ee
which is also the case in a relativistic treatment \cite{K94}. For a strong shock, $B_2\to4B_1$, the ratio (\ref{maxratio}) is $(2+\sqrt{3})/2-\sqrt{3}=13.93$. However, the energy ratio decreases rapidly away from the point in the momentum space of the incident particles where this maximum occurs \cite{BM01}. 

\begin{figure}[t]
\centerline{
\psfig{figure=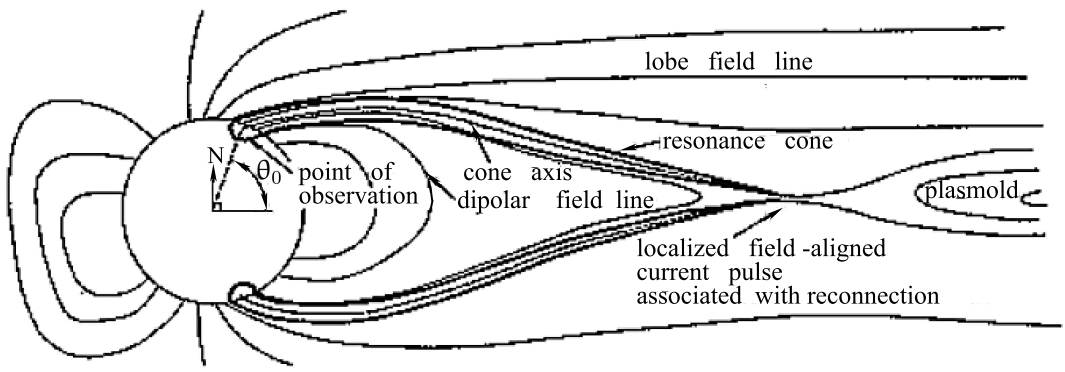,height=50mm}}
\caption{The foreshock may be separated
into an electron foreshock, nearer to the field line
that is tangent to the bow shock, and an ion foreshock,
nearer to the bow shock; the dashed line indicates the
ill-defined boundary between these two regions, which
actually merge continuously into each other.}
\label{fig:acc6}
\end{figure}

SDA accounts well for relevant observations of particles associated with shocks in the heliosphere, in particular, the Earth's bow shock and other planetary bow shocks. SDA produces a foreshock region ahead of any (convex) curved shock. The foreshock is the region ahead of the shock and behind the field line that is tangent to the shock, as illustrated in figure~\ref{fig:acc6}. Fast particles reflected from the shock propagate along the field lines as they are swept towards the shock by the upstream flow. Electrons have higher speeds than ions, so that they propagate further before encountering the shock, producing an electron foreshock region that is much larger than the ion foreshock region. SDA should similarly populate a foreshock region ahead of any curved shock.

Type~II solar radio bursts are directly associated with shocks in the solar corona, and extending into the solar wind \cite{NelsonM85}. Acceleration by SDA is a favored mechanism for the acceleration of electrons that produce type~II radio bursts \cite{HolmanPesses83,Knock}. SDA is also a potential candidate for preacceleration for DSA. Although both suggestions seem plausible, additional ideas are required to account for the details of the observations.

One modification of DSA, called shock surfing acceleration \cite{UcerS01}, involves the particles being trapped in the shock front for much longer than the simple theory of SDA implies. In other modifications, the assumption that the shock is a simple discontinuity is relaxed. Various additional possible acceleration mechanisms arise: due to a potential difference along field lines \cite{JE87}; various instabilities that can lead to large-amplitude waves that can trap and preaccelerate particles (`surfatron' acceleration \cite{McClements01}); and nonstationarity that involves the shock continually breaking up and reforming \cite{SCD02}.

\subsection{Resonant Acceleration}

MHD turbulence consists of a mixture of the Alfv\'en and fast modes. Fermi-type stochastic acceleration involves only the fast mode, and may be interpreted in terms of Cerenkov resonance, at harmonic number $s=0$. Resonances at $s\ne0$ are possible and lead to dissipation of turbulence in both modes, with the energy going into the particles. For highly relativistic particles, resonances at high harmonics lead to a diffusion in momentum space, somewhat analogous to Fermi-type acceleration, but with a diffusion coefficient that depends on a higher power of $p$ \cite{Lacombe77}. Although this acceleration must occur, for say GCRs and turbulence in the interstellar medium, there is no strong case for it playing an important role in specific applications.

\subsection{Acceleration during magnetic reconnection}

Magnetic energy release through magnetic reconnection is known to be associated with powerful events, notably magnetospheric substorms and solar flares, that lead to acceleration of fast particles. There have been various suggestions that reconnection can lead directly to acceleration of fast particles. For example, for magnetic reconnection in a neutral sheet, where the magnetic field is zero on a surface and ${\bi B}$ changes sign across this surface, a test particle that enters the neutral sheet typically emerges with a higher energy \cite{Speiser65}. Although such energy gains occur, they do not appear to constitute an important acceleration mechanism. Both theory and observation suggest that the acceleration is an indirect, rather than a direct consequence of the reconnection. 

A different application of heating in current sheets has been proposed in connection with pulsar winds \cite{C90,LK01}.  Beyond the light cylinder there is a wind containing wound up magnetic field lines in a nearly azimuthal direction, with the sign of ${\bi B}$ reversing periodically with radial distance. The wind accelerates particles at a termination shock, where it encounters the surrounding synchrotron nebula. An outstanding problem is that the energy flux is Poynting dominated near the light cylinder, but must be kinetic-energy dominated before it reaches the termination shock. The suggestion is that magnetic energy dissipation in the neutral sheets separating the regions of opposite  ${\bi B}$ provide the necessary conversion of magnetic energy into particle energy \cite{C90,LK01}. An alternative suggestion is that the number of current carriers becomes inadequate leading to the development of an accelerating field \cite{MM96}.

\section{Future directions}

Three generic acceleration mechanisms are widely recognized, and there are also a number of other specific mechanisms that are important in specific applications. The pioneering works of Fermi \cite{F49,F54} anticipated two of three generic mechanisms: stochastic acceleration and diffusive shock acceleration (DSA), respectively. The third generic mechanism, acceleration by a parallel electric field, $E_\parallel$, is the least adequately understood. 

The most important acceleration mechanism in astrophysics is DSA. It is the accepted mechanism for the acceleration of GCRs and for relativistic electrons in most synchrotron sources. The favored location for the acceleration of GCRs is at the shocks in young supernova remnants \cite{Hillas06}. Despite the basic theory being well established there are essential details where our current understanding is incomplete and where further progress is to be expected. The well-established theory is a test-particle model for DSA at a single nonrelativistic shock. Several modifications to the theory are known to be important: the generalizations to relativistic shocks, and to multiple shocks, and the dynamical effects of the accelerated particles on the shock structure.  There are several aspects of the required efficient scattering in DSA that are inadequately understood. One is the injection problem: resonant scattering of nonrelativistic ions and electrons involves Alfv\'en and whistler waves, and these require that the particles already have speeds $v>v_A$ and $v>43v_A$, respectively, and preacceleration to above these thresholds is required before DSA can operate. For ions, the elemental abundances of GCRs suggests an injection mechanism that is insensitive to ionic species, whereas for flare-associated solar energetic particles there are extreme elemental anomalies (notably $^3$He). For electrons the injection problem is more severe, and less understood. Second is the long-standing question of the ratio of relativistic electrons to ions \cite{GS64}. Unfortunately, in situ data on shocks in the heliosphere provide little guidance on this problem: evidently the shocks in synchrotron sources have much higher Mach numbers than those in the heliosphere, for which there is little evidence for DSA producing relativistic electrons. A third aspect relates to resonant scattering of the highest energy particles. Unlike lower-energy particles, the resonant waves needed to scatter higher-energy particles cannot be generated by the particles themselves, and other sources of the required waves are speculative. A suggestion that may overcome this difficulty is that the lower-energy particles cause waves to grow through a nonresonant instability, which leads to amplification of the magnetic field by a very large factor \cite{Bell04}, and increases the maximum energy to which DSA can operate. Fourth, spatial diffusion across field lines is postulated to occur in DSA at perpendicular shocks, but is incompletely understood \cite{AB94}. 

Despite its apparent simplicity, acceleration by $E_\parallel$ is the least understood of the three generic mechanisms. An important qualitative point is that $E_\parallel\ne0$ is inconsistent with MHD, and acceleration by $E_\parallel$ necessarily involves concepts that are relatively unfamiliar in the context of MHD. In particular, one requires a driver that sets up an EMF, or available potential, and this driver is typically far removed from the acceleration region where the potential partially localizes to produce the accelerating $E_\parallel$. The coupling between these is best understood application is to acceleration of auroral electrons associated with magnetospheric substorms, but  how the $E_\parallel$ develops remains uncertain. Existing models for acceleration by $E_\parallel$ in `gaps' in pulsars and in bulk energization of electrons in solar flares have serious flaws. The present difficulties may be solved by assuming that the relevant $E_\parallel$ is oscillating, associated with inertial Alfv\'en waves in an Alfv\'en resonator in the magnetosphere, and with large-amplitude oscillations in pulsar magnetospheres. Further progress in understanding acceleration by $E_\parallel$ is to be expected.

Compared with the rapid evolution in ideas in other subfields of astrophysics, the ideas involved in understanding particle acceleration have developed only gradually over many decades. Further progress is certain to occur, but is likely to involve gradual acceptance of some of the existing ideas, rather than radical new insights.

\appendix
\section{Quasilinear equations}

The quasilinear equations are written down here using a semiclassical formalism. The waves are regarded as a collection of wave quanta, with energy $\hbar\omega$ and momentum $\hbar{\bf k}$. Waves in an arbitrary mode, labeled $M$, have dispersion relation $\omega=\omega_M({\bf k})$, polarization vector ${\bf e}_M({\bf k})$ and ratio of electric to total energy, $R_M({\bf k})$. Their distribution is described by their occupation number $N_M({\bf k})$. A resonant interaction occurs when the gyroresonance condition is satisfied:
\be
\omega -s\Omega-k_\parallel v_\parallel=0,
\label{(8.7)}
\ee

\noindent where $s$ is an integer, $\Omega=|q|B/m\gamma$ is the relativistic gyrofrequency, and $k_\parallel$, $v_\parallel$ are the components of ${\bf k}$, ${\bf v}$ parallel to ${\bf B}$. Resonances at $s>0$ are said to be via the normal Doppler effect, and those at $s<0$ are said to be via the  anomalous
Doppler effect. Resonances at $s\le0$ are possible only for waves with refractive index greater than unity. The effect of a wave-particle interaction is described by the probability of spontaneous emission, $w_M({\bf p},{\bf k},s)$, which is the probability per unit time that the particle emit a wave quantum in the wave mode $M$ in the elemental range $d^3{\bf k}/(2\pi)^3$. For a particle of charge $q$ and mass $m$,  the probability of spontaneous emission is given by
\bea
&&w_M({\bf p},{\bf k},s)=
{2\pi q^2R_M({\bf k})\over
\varepsilon_0\hbar\omega_M({\bf k})}\, 
\big|{\bf e}_M^*({\bf k})\cdot
{\bf V}({\bf k},{\bf p};s)\big|^2\,
\delta\big(\omega_M({\bf k}) 
-s\Omega-k_\parallel v_\parallel\big),
\nn
\\
\ms
&&\qquad
{\bf V}({\bf k},{\bf p};s)=
\big(
v_\perp{s\over k_\perp R}J_s(k_\perp R),
-i\eta v_\perp J'_s(k_\perp R),
v_\parallel J_s(k_\perp R)
\big),
\nn
\\
\ms
&&\qquad\qquad
\Omega={\Omega_0\over\gamma},
\quad
\Omega_0={|q|B\over m},
\quad
R={v_\perp\over\Omega}={p_\perp\over|q|B}.
\label{(8.12)}
\eea

An advantage of the semiclassical formalism is that the Einstein coefficients imply that the probability of stimulated emission and true absorption are given by multiplying this probability by $w_M({\bf p},{\bf k},s)N_M({\bf k})$. This allows the effect on the occupation numbers of the waves and the particles to be written down using a simple bookkeeping argument. For the waves one finds
\be
{dN_M({\bf k})\over dt}=\left(
{dN_M({\bf k})\over dt}
\right)_{\rm spont}
-\gamma_M({\bf k})N_M({\bf k}),
\label{(8.16)}
\ee
where the two terms on the right hand side describe spontaneous emission and absorption, respectively, with the absorption coefficient given by
$$
\gamma_M({\bf k})
=-\sum_{s=-\infty}^\infty
\int d^3{\bf p}\,
w_M({\bf p},{\bf k},s)
{\hat D}_sf({\bf p}),
$$
\be
{\hat D}_s=\hbar\left(
{s\Omega\over v_\perp}{\pd\over\pd p_\perp}
+k_\parallel{\pd\over\pd p_\parallel}
\right)=
{\hbar\omega\over v}
\left(
{\pd\over\pd p}
+{\cos\alpha-n_M\beta\cos\theta
\over p\sin\alpha}
{\pd\over\pd \alpha}
\right).
\label{(8.17)}
\ee
The two forms correspond to, respectively, cylindrical and polar coordinates in momentum space. 

The evolution of the distribution of particles due to the resonant interaction includes a term, neglected here, that describes the effect of spontaneous emission, and a quasilinear diffusion equation that describes the effect of the induced processes. In cylindrical and spherical polar coordinates, this equation is
\bea
{d f({\bf p})\over d t}
&=&{1\over p_\perp}{\pd\over \pd p_\perp}
\left\{p_\perp \left[D_{\perp\perp}({\bf p})
{\pd\over\pd p_\perp}
+D_{\perp\parallel}({\bf p})
{\pd\over\pd p_\parallel}
\right]
f({\bf p})\right\}
\nn
\\
&&\qquad\qquad\qquad
+{\pd\over \pd p_\parallel}
\left\{ \left[D_{\parallel\perp}({\bf p})
{\pd\over\pd p_\perp}
+D_{\parallel\parallel}({\bf p})
{\pd\over\pd p_\parallel}
\right]
f({\bf p})\right\}
\nn
\\
&=&{1\over\sin\alpha}{\pd\over \pd\alpha}
\left\{\sin\alpha \left[D_{\alpha\alpha}({\bf p})
{\pd\over\pd\alpha}
+D_{\alpha p}({\bf p})
{\pd\over\pd p}
\right]
f({\bf p})\right\}
\nn
\\
&&\qquad\qquad\quad
+{1\over p^2}{\pd\over \pd p}
\left\{ p^2\left[D_{p\alpha}({\bf p})
{\pd\over\pd\alpha}
+D_{pp}({\bf p})
{\pd\over\pd p}
\right]
f({\bf p})\right\},
\qquad\qquad
\label{(8.18)}
\eea
respectively, with the diffusion coefficients in either set of coordinates given by
$$
D_{QQ'}({\bf p})
=\sum_{s=-\infty}^\infty
\int{d^3{\bf k}\over(2\pi)^3}\,
w_M({\bf p},{\bf k},s)\,
\Delta Q\,\Delta Q'\,N_M({\bf k}),
\qquad
\Delta Q={\hat D}_sQ,
$$
\be
\Delta  p_\perp={s\Omega\over v_\perp},
\quad
\Delta  p_\parallel=\hbar k_\parallel,
\quad
\Delta\alpha={\hbar(\omega\cos\alpha-k_\parallel v)\over
pv\sin\alpha},
\quad
\Delta  p={\hbar\omega\over v}.
\label{(8.19)}
\ee

\section*{Further reading: monographs and reviews}

\noindent
Alfv\'en H F\"althammar C-G
(1963)
{Cosmical Electrodynamics}.
Oxford University Press

\noindent
Aschwanden MJ
(2004)
Physics of the solar corona.
Springer, Berlin

\noindent
Benz AO
(1993)
Plasma astrophysics: kinetic processes in solar and stellar coronae. Kluwer Academic Publishers, Dordrecht

\noindent
Berezinskii VS, Bulanov SV, Dogiel VA, Ginzburg VL, Ptuskin VS 
(1990) Astrophysics of cosmic rays. North Holland, Amsterdam

\noindent
Dorman LI
(2006)
Cosmic ray interactions, propagation, and acceleration in space plasmas.
Springer, New York 

\noindent
Drury LO'C (1983)
An introduction to the theory of diffusive shock acceleration of energetic particles in tenuous plasmas,
Rep Prog Phys 46:973--1027

\noindent
McLean DJ, Labrum NR (Eds) (1986)  Solar
Radiophysics, Cambridge University Press

\noindent
Melrose DB
(1980)
Plasma Astrophysics  Volume I \& II.
Gordon \& Breach, New York

\noindent
Melrose DB
(1986)
{Instabilities in Space and Laboratory Plasmas}.
Cambridge University Press

\noindent
Malkov MA, Drury LO'C (2001) Nonlinear theory of diffusive acceleration of particles by shock waves. Rep Prog Phys 64:429--481

\noindent
Michel FC
(1991)
Physics of neutron star magnetospheres.
The University of Chicago Press

\noindent
Priest ER, Forbes T
(2000)
Magnetic reconnection --- MHD theory and applications.
Cambridge University Press

\noindent
Schlickeiser R (2002) Cosmic ray astrophysics. Springer, Berlin

\noindent
Stone RG, Tsurutani BT (Eds)
(1985)
Collisionless shocks in the heliosphere: a tutorial review.
Geophysicial Monograph 34, 
American Geophysical Union, Washington DC

\noindent
Sturrock PA
(1980)
Solar flares.
Colorado Associated University Press

\noindent
\u Svestka Z
{1976}
{Solar Flares}.
{D Reidel, Dordrecht}

\noindent
Vogt J
(2002)
Alfv\'en wave coupling in the auroral current circuit.
Surveys Geophys 23:335--377

\end{document}